\documentclass[longbibliography,groupedaddress,showpacs,showkeys,amssymb,eqsecnum,aps,nofootinbib,superscriptaddress]{revtex4}
\usepackage[pdftex]{graphicx}
\usepackage[pdftex]{graphics}
\usepackage{amsmath} 
\usepackage{epsf}                                                                                           
\usepackage{color}                   
\usepackage{verbatim}                                                                         
\usepackage{hyperref}

\newcommand{\be}{\begin{equation}}

\newcommand{\dc}{{\cal D}}
\newcommand{\lc}{{\cal L}}

\newcommand{\ee}{\end{equation}}

\begin{document}                                                                                              


\title{Use of Lagrange multiplier fields to eliminate multiloop corrections}
\author{F. T. Brandt}  
\email{fbrandt@usp.br}
\affiliation{Instituto de F\'{\i}sica, Universidade de S\~ao Paulo, S\~ao Paulo, SP 05508-090, Brazil}

\author{J. Frenkel}
\email{jfrenkel@if.usp.br}
\affiliation{Instituto de F\'{\i}sica, Universidade de S\~ao Paulo, S\~ao Paulo, SP 05508-090, Brazil}

\author{D. G. C. McKeon}
\email{dgmckeo2@uwo.ca}
\affiliation{
Department of Applied Mathematics, The University of Western Ontario, London, ON N6A 5B7, Canada}
\affiliation{Department of Mathematics and Computer Science, Algoma University,
Sault Ste.~Marie, ON P6A 2G4, Canada}

\author{G. S. S. Sakoda}    
\email{gustavo.sakoda@usp.br}
\affiliation{Instituto de F\'{\i}sica, Universidade de S\~ao Paulo, S\~ao Paulo, SP 05508-090, Brazil}

\date{\today}

\begin{abstract}
The problem of eliminating divergences arising in quantum gravity is
generally addressed by modifying the classical Einstein-Hilbert
action. These modifications might involve the introduction of local
supersymmetry, the addition of terms that are higher-order in the
curvature to the action, or invoking compactification of superstring
theory from ten to four dimensions. An alternative to these approaches
is to introduce a Lagrange multiplier field that restricts the path
integral to field configurations that satisfy the classical equations
of motion; this has the effect of doubling the usual one-loop
contributions and of eliminating all effects beyond one loop. We show
how this reduction of loop contributions occurs and find the gauge
invariances present when such a Lagrange multiplier is introduced into
the Yang-Mills and Einstein-Hilbert actions. Moreover, we quantize
using the path integral, discuss the renormalization, and then show
how Becchi-Rouet-Stora-Tyutin (BRST) invariance can be used to both
demonstrate that unitarity is retained and to find BRST relations
between Green’s functions. In the Appendices we show how background
field quantization can be implemented, consider the use of a Lagrange
multiplier field to restrict higher-order contributions in
supersymmetric theories, and derive the BRST equations satisfied by
the generating functional.
\end{abstract}                                                                                                


\maketitle

\section{Introduction}   
The removal of ultraviolet divergences that arise in quantum field theory has been a long-standing problem.  The issue was treated systematically by Dyson \cite{Dyson:1949bp}, but even then divergences arising beyond one-loop order presented special problems \cite{PhysRev.82.217,Collins:105730}. Renormalization to all orders is feasible only if the coupling constant is dimensionless unless symmetries that are present in the classical Lagrangian result in a cancellation between different divergent contributions. This is what makes renormalization of divergences arising when one quantizes the Einstein-Hilbert (EH) action problematic. At one-loop order, divergences are proportional to terms that vanish if the classical equation of motion for the background field are satisfied and surface terms are discarded, so a shift in the background field eliminates these divergences \cite{tHooft:1974bx,tHooft:2002xp,Buchbinder:1992rb},
but beyond one-loop order \cite{Goroff:1986th,vandeVen:1992gw}
or if the metric couples to matter fields
\cite{tHooft:1974bx,vandeVen:1992gw,Deser:1974xq,Deser:1974cy}
this shift is no longer feasible. This makes it apparent that quantization and renormalization of the gravitational field are necessarily different from the procedures that worked in electrodynamics and Yang-Mills (YM) theory.

A great deal of effort has been devoted to modifying the gravitational
theory to obtain a model that is renormalizable and unitary while
having General Relativity (GR) as a classical limit when in four dimensions. These attempts have met with varying degrees of success. In supergravity models, the usual generators of Poincar\'e symmetry are supplemented with Fermionic generators of a local gauge
transformation \cite{VanNieuwenhuizen:1981ae,Freedman:2012zz,weinberg:book2000}.
The resulting supergravity theories have improved ultraviolet behaviour  \cite{PhysRevLett.38.527},
but even the most symmetric of the models (N=8 supergravity) may or may not be
renormalizable \cite{PhysRevD.98.086021},
although at lower orders in the loop expansion some unexpected cancellations
of divergences appear \cite{Stelle:2011zeu}.
In these supergravity models, there are also superparticles that have as yet been unobserved.


Another approach is to introduce into the EH action terms of quadratic
or higher-order in the Riemann tensor. These terms suppress the
contribution of the graviton propagator and hence can make it possible to
have renormalization \cite{Stelle:1976gc}. However, these propagators may now
have unacceptable ghost poles, though it is possible that this
shortcoming may be overcome \cite{Mannheim:2011ds}. Besides, it may not be
possible to reconcile these extra terms in the classical action with observations.

Treating the EH action as a low energy limit of a superstring is also a possible approach to the problem of quantizing gravity. This approach has many appealing features, but it also gives rise to several
problems \cite{green_schwarz_witten_2012,polchinski_1998,west_2012}.
The first quantized version of the model requires it to be defined in ten dimensions, which must be compactified in some way to $3+1$ dimensions. There is no fully satisfactory way of second quantizing the superstring. There is a myriad of ground states. Many as yet unobserved particles are predicted. Incorporating the Standard Model into superstring theory has not proved to be straightforward.


Loop quantum gravity \cite{Gambini:2011zz} is another approach to
quantizing gravity, but it too faces problems when reconciling with
GR when treating a dynamical metric coupled to matter fields in the classical limit.

It has also been proposed that couplings induced by divergences that arise
in the quantized EH action are ``asymptotically safe'' \cite{Nagy:2012ef}; that is, they approach a fixed point in the high energy limit. This has proved to be difficult to demonstrate.

If one were to simply do a canonical analysis of the EH action \cite{Dirac:1958sc,Arnowitt:1962hi}
then it is difficult to obtain covariant results when using the constraint structure to quantize the theory.

We propose a relatively uncomplicated way of quantizing the EH action
that makes it possible to remove divergences induced by quantum
effects without losing unitarity and which retains GR
in the classical limit. It involves simply introducing a term into the
action in which a Lagrange multiplier (LM) field is used to ensure
that the classical equations of motion are satisfied. If one quantizes
a classical Lagrangian for a field $\phi_i$ that has been supplemented
by such a term by using the path integral formalism, it can be shown
that the one-loop radiative corrections to the classical action are
twice those that occur if the term involving the LM were absent, and
that all radiative effects beyond one-loop order are absent. The
introduction of such an LM field has been considered in 
YM theory \cite{McKeon:1992rq}
in the Proca model \cite{Chishtie:2012hn} 
and with the EH Lagrangian \cite{Brandt:2018lbe}.

Not having any radiative effects arising beyond one-loop order simplifies the renormalization procedure needed to remove ultraviolet divergences that may arise. This is of particular relevance in dealing with the EH action; we will discuss below how all divergences (confined to one-loop order with the introduction of the LM field) can be absorbed into the LM field even when the metric interacts with matter fields.

There are precedents for considering actions in which a field occurs linearly, much like the LM field we will use.  For example, Jackiw \cite{Jackiw:1984je} and Teitelboim \cite{Teitelboim:1983uyTeitelboim:1983ux} have provided dynamics for the metric field in $1+1$ dimensions by introducing a field that occurs linearly in the action;
Chamseddine \cite{Chamseddine:1990hr}
has used this approach in conjunction with the Bosonic string.
In another example, Gozzi \cite{Gozzi:1986ge} has considered
a path integral formulation of classical mechanics by using a LM
to completely reduce the path integral to a Dirac delta function which restricts all field configuration
to those satisfying the classical equations of motion.
Witten \cite{Witten:1988hc} has discussed the Einstein-Cartan form of the action
in GR in $2+1$ dimensions; in this action the ``dreibein'' field occurs only linearly,
making it possible to solve the theory. Having a fields occur only linearly
in the action is also a feature of the Palatini form of the EH action in $1+1$
dimensions \cite{McKeon:2005be}.


In the next section, we show how the introduction of  a LM can be used
to eliminate all loop diagrams beyond one-loop order and to double the
usual one-loop contribution.  This is done for an arbitrary theory,
quantized using the path integral. We present both a general argument
and one which employs Feynman diagrams. Next, a discussion of how the
gauge symmetries, present in the original action, are affected by the
introduction of the LM field and the consequences for quantization
examined when using the Faddeev-Popov (FP) procedure \cite{Faddeev:1967fc}.

We then apply these general considerations to the YM and EH actions. How the surviving one-loop divergences can be removed is then discussed.

An analysis of the BRST \cite{brs74,Tyutin:1975qk} invariance of both the YM and EH actions is then considered.
This approach can be used both to relate different
Green's functions and to prove unitarity in these models.

In the first appendix, we show how the use of a background field \cite{Dewitt:1967ub, Abbott:1980hwabbott82}
when quantizing modifies the LM approach. We do this as performing perturbative calculations with the EH action is only feasible if a background field is introduced.
By adopting the approach of  Refs. \cite{Abbott:1980hwabbott82},
we show that using the background field should not be viewed as a deficiency of a quantization procedure for GR as is sometimes stated. The background field is naturally introduced by the formalism; it is not an ad hoc entity that leaves the resulting quantized theory somehow non-covariant. 

The second appendix deals with two supersymmetric models. First, the Wess-Zumino model with the LM field is considered, and then supergravity is discussed.
The BRST equations are derived in a third appendix.

\section{The general formalism}
Let us consider the action
\be\label{eq1}
S[\phi_i] = \int dx\lc(\phi_i)
\ee
for a field $\phi_i$. 
The path integral quantization procedure makes use of the generating functional 
\be\label{eq2}
Z[j_i] = \int\dc\phi_i \exp \frac{i}{\hbar} \int dx (\lc(\phi_i) + j_i\phi_i).
\ee
In Eq. \eqref{eq2} we now make the replacement \cite{Brandt:2018lbe}
\be\label{eq3}
\lc(\phi_i) + j_i\phi_i \rightarrow\lim_{\eta\rightarrow\infty}
\left[\frac{1+\eta}{2}\left(\lc({\phi_+}_i)+j_i{\phi_+}_i\right)+
\frac{1-\eta}{2}\left(\lc({\phi_-}_i)+j_i{\phi_-}_i\right)\right],
\ee
where
\be\label{eq3a}
{\phi_{\pm}}_i \equiv \phi_i \pm \frac{1}{\eta} \lambda_i,
\ee
then Eq. \eqref{eq2} becomes (if we set $\hbar=1$)
\be\label{eq4}
Z[j_i] = \int\dc\lambda_i\dc\phi_i \exp {i} \int dx \left[\lc(\phi_i) 
+\lambda_j\frac{\partial\lc(\phi_i)}{\partial\phi_j}
+ j_i(\phi_i+\lambda_i)\right].
\ee
Integration over the LM field $\lambda_i$ results in a functional 
$\delta$-function so that
\be\label{eq5}
Z[j_i] = \int\dc\phi_i \delta\left[
\frac{\partial\lc}{\partial\phi_j} + j_j
\right]
\exp {i} \int dx \left[\lc(\phi_i) + j_i \phi_i\right].
\ee
One can now make use of the functional analogue of the integral 
\cite{McKeon:1992rq,Chishtie:2012hn,Brandt:2018lbe}
\be\label{eq6}
\int_{-\infty}^{\infty} dx \delta(f(x)) g(x) = \sum_i\frac{g(\bar x_i)}
{|f^\prime(\bar x_i)|}
\ee
where ${\bar x}_i$ is a solution of 
\be\label{eq7}
f(\bar x_i) = 0.
\ee
It is the functional analogue of $|f^\prime({\bar x}_i)|$ occurring in
\eqref{eq6} that leads to a functional determinant appearing once the
functional integral over $\phi_i$ has been performed in Eq. \eqref{eq5};
we obtain
\be\label{eq8}
Z[j_i] = \sum_i \frac{\exp i \int dx\left[\lc(\bar\phi_i)+j_i\bar\phi_i\right]}
{\det\left(\frac{\partial^2\lc(\bar\phi_i)}{\partial\bar\phi_j\partial\bar\phi_k}\right)},
\ee
where $\bar\phi_i$ is a solution to the classical equation
\be\label{eq9}
\frac{\partial{\cal L}(\bar\phi_i)}{\partial\bar\phi_j}+j_j=0.
\ee
The exponential in Eq. \eqref{eq8} is the result of summing all tree
  diagrams \cite{Boulware:1968zz} while the functional determinant is
  the square of all one-loop diagrams that arise from Eq. \eqref{eq4}
  if $\lambda_i$ were absent (ie the action of Eq. \eqref{eq2} alone
  were used). No contributions beyond one-loop order arise. External
  fields are on shell.

This general result can be recovered using an argument that employs
Feynman diagrams. If we expand $\lc(\phi_i)$ so that
\be\label{eq10}
\lc(\phi_i) = \frac{1}{2!} a_{ij}^{(2)} \phi_i\phi_j
+\frac{1}{3!} a_{ijk}^{(3)} \phi_i\phi_j\phi_k
+\frac{1}{4!} a_{ijkl}^{(4)} \phi_i\phi_j\phi_k\phi_l+\cdots
\ee
then in Eq. \eqref{eq4} we find that (with $a^{(n)}$ symmetric)
\begin{eqnarray}\label{eq11}
&&\lc +\lambda_j\frac{\partial\lc}{\partial\phi_j} + j_i(\phi_i+\lambda_i)
\nonumber\\ &=&
\sum_{n=2}\left(\frac{1}{n!}a_{i_1\cdots i_n}^{(n)} \phi_{i_1}\cdots \phi_{i_n}
+\frac{1}{(n-1)!}a_{i i_2\cdots i_n}^{(n)} \lambda_i \phi_{i_2}\cdots \phi_{i_n}
\right) + j_i(\phi_i+\lambda_i).
\end{eqnarray}
The terms bilinear in the fields $\lambda_i$, $\phi_i$ in
Eq. \eqref{eq11} that give rise to the propagators are
\be\label{eq12}
\lc^{(2)} =
\frac{1}{2!}\left(\begin{array}{ll}\phi_i,\lambda_i\end{array}\right)
\left(\begin{array}{cc} a^{(2)}_{ij} & a^{(2)}_{ij} \\ 
                               a^{(2)} _{ij} & 0\end{array}\right)
\left(\begin{array}{c}\phi_j \\ \lambda_j\end{array}\right).
\ee
Since 
\be\label{eq12a}
\left(\begin{array}{cc} a & a \\
                                a  & 0\end{array}\right)^{-1} =
\left(\begin{array}{cc} 0       & a^{-1} \\
                                a^{-1}         & -a^{-1}  \end{array}\right)
\ee
we see that there is no propagator $\langle\phi_i \phi_j\rangle$ for
the field $\phi_i$. There are, however, mixed propagators 
$\langle \lambda_i\phi_j\rangle = \langle \phi_i\lambda_j\rangle$ and
a propagator for the $\langle \lambda_i\lambda_j\rangle$
for the LM field; all the vertices have at most a single LM field attached
to $n\ge 2$ $\phi_i$ fields.
As a result, one cannot draw a Feynman diagram with
more than one loop and only the $\phi_i$ field can appear on an
external leg. Furthermore, only mixed propagators are allowed in the
internal lines.
For example, the only contributions to the four point function that follow from
the Feynman rules derived from Eq. \eqref{eq11},
are only one-loop order, and they are shown in Fig. 1. No higher loop
diagrams contribute.

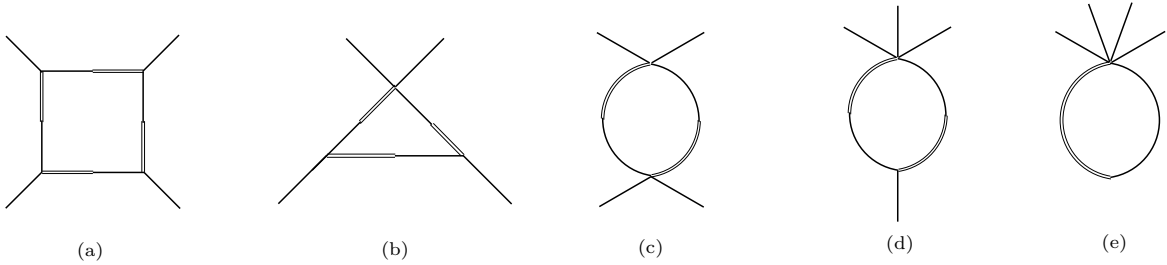
\begin{figure}[h]
\begin{picture}(0,0)%
\includegraphics{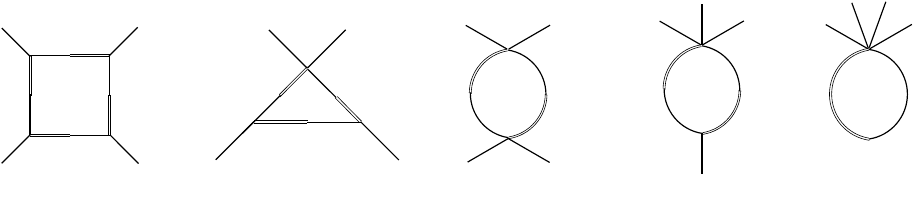}%
\end{picture}%
\setlength{\unitlength}{2486sp}%
\begingroup\makeatletter\ifx\SetFigFont\undefined%
\gdef\SetFigFont#1#2#3#4#5{%
  \reset@font\fontsize{#1}{#2pt}%
  \fontfamily{#3}\fontseries{#4}\fontshape{#5}%
  \selectfont}%
\fi\endgroup%
\begin{picture}(11603,2619)(-6905,-6392)
\put(4044,-6241){\makebox(0,0)[lb]{\smash{{\SetFigFont{7}{8.4}{\rmdefault}{\mddefault}{\updefault}{\color[rgb]{0,0,0}(e)}%
}}}}
\put(-3142,-6309){\makebox(0,0)[lb]{\smash{{\SetFigFont{7}{8.4}{\rmdefault}{\mddefault}{\updefault}{\color[rgb]{0,0,0}(b)}%
}}}}
\put(-6176,-6328){\makebox(0,0)[lb]{\smash{{\SetFigFont{7}{8.4}{\rmdefault}{\mddefault}{\updefault}{\color[rgb]{0,0,0}(a)}%
}}}}
\put(-582,-6293){\makebox(0,0)[lb]{\smash{{\SetFigFont{7}{8.4}{\rmdefault}{\mddefault}{\updefault}{\color[rgb]{0,0,0}(c)}%
}}}}
\put(1892,-6250){\makebox(0,0)[lb]{\smash{{\SetFigFont{7}{8.4}{\rmdefault}{\mddefault}{\updefault}{\color[rgb]{0,0,0}(d)}%
}}}}
\end{picture}%
\caption{One-loop contributions to
  $\langle\phi_i\phi_j\phi_k\phi_l\rangle$. Internal lines contain only
  mixed propagators and the vertices contain a single LM field}
\end{figure}

It can be shown that the combinatoric factor for such one-loop
diagrams is always twice the one associated with the usual Feynman
diagrams, when there is no LM field. As a simple example of this
property, one can compare the combinatorial factors of each diagram in Fig. 1, with the
corresponding ones obtained from the theory described by the action in
Eq. \eqref{eq1}. Consequently, the Feynman diagrams that follow from
Eq. \eqref{eq11} give results consistent with the of Eq. \eqref{eq8}.

In gauge theory models, $a^{(2)}_{ij}$ has no inverse as it has
vanishing eigenvalues; in such models the action $S$ in 
Eq. \eqref{eq1} is form invariant under the replacement
\be\label{eq13}
\phi_i\rightarrow \phi^\prime_i = \phi_i + H_{ij}(\phi_k) \xi_j. 
\ee
In order to render the path integral of Eq. \eqref{eq4} well defined 
when this happens, we adopt the FP 
procedure. First though we note that if the action $S$ in 
Eq. \eqref{eq1} is invariant under \eqref{eq13}, then 
\be\label{eq14}
\int dx \lc(\phi^\prime_i) = \int dx \left(\lc(\phi_i)+H_{ij}(\phi_i)\xi_j 
\frac{\partial\lc}{\partial\phi_i} \right)= \int dx\lc(\phi_i) 
\ee
and so by Eq. \eqref{eq14} we have form invariance if
\be\label{eq15}
\lambda_i\rightarrow \lambda^\prime_i = \lambda_i + H_{ij}(\phi_k) \zeta_j. 
\ee
In addition, if under the gauge transformation of Eq. \eqref{eq13},
\be\label{eq16}
\int dx \left(\lc(\phi_k) +\lambda_i \frac{\partial\lc(\phi_k)}{\partial\phi_i} \right)
= \int dx \left(\lc(\phi_k^\prime) +\lambda_i \frac{\partial\phi^\prime_j}{\partial\phi_i}
\frac{\partial\lc(\phi^\prime_k)}{\partial\phi^\prime_j} \right)
\ee
there is form invariance if 
\be\label{eq17}
\lambda^\prime_i = \lambda_i + \lambda_l \frac{\partial
  H_{ij}(\phi_k)}{\partial\phi_l}  \xi_j. 
\ee
We thus see that there are now two gauge invariances associated with
the action appearing in Eq. \eqref{eq14}; there is that of
Eqs. \eqref{eq13} and \eqref{eq17} as well as that of
Eq. \eqref{eq15}.

In order to eliminate the degeneracy occurring in the path integral of
Eq. \eqref{eq4} due to these two gauge invariances, we adopt the usual
FP procedure \cite{Faddeev:1967fc}. 

We choose to use the same gauge condition for the fields $\phi_i$ and
$\lambda_i$
\be\label{eq18}
F_{ij} \phi_j = 0 = F_{ij} \lambda_j .
\ee
These need not be the same; indeed it is possible to have more than
one gauge condition applied to a gauge field
\cite{Brandt:2007td}. Next, a constant factor
\begin{eqnarray}\label{eq19}
\int \dc\xi_i\dc\zeta_i\delta\left[
F_{ij}\left(\left(\begin{array}{c} \phi_j \\
         \lambda_j\end{array}\right)+
\left(\begin{array}{cc} 0 & H_{jk} \\
        H_{jk} & \lambda_l \frac{\partial H_{jk}}{\partial\phi_l}\end{array}\right)
\left(\begin{array}{c} \zeta_k \\         \xi_k\end{array}\right)\right)-
\left(\begin{array}{c} p_i \\        q_i\end{array}\right)
\right]
\det 
\left(\begin{array}{cc} 0 & F_{ij} H_{jk} \\
        F_{ij} H_{jk} & F_{ij} \left(\lambda_l \frac{\partial H_{jk}}{\partial\phi_l}\right)\end{array}\right)
\end{eqnarray}
is inserted into the path integral of Eq. \eqref{eq4}. The functional
determinant in Eq. \eqref{eq19} can be rewritten in several useful
ways, as
\begin{subequations}\label{eq20}
\be\label{eq20a}
\det\left(\begin{array}{cc} 0 & A \\ A & B \end{array}\right)
=-{\det}^2 A
\ee
or
\be\label{eq20b}
=\det\left(\begin{array}{cc} 0 & A \\ A & A+B \end{array}\right),
\ee
\end{subequations}
with $A$ and $B$ identified with $F_{ij} H_{jk}$ and
$F_{ij} \left(\lambda_l \frac{\partial H_{jk}}{\partial\phi_l}\right)$
respectively. The usual FP determinant when there is no LM is
$\det A$, so the result of Eq. \eqref{eq20a} is expected. When we
discuss the BRST invariance below, Eq. \eqref{eq20b} is more useful;
we introduce Fermionic ghost fields $c_i$, $\bar c_i$, $d_i$, so that
\be\label{eq21}
\det\left(\begin{array}{cc} 0 & A \\ A & A+B \end{array}\right)
=\int \dc c_i \dc \bar c_i \dc d_i \dc \bar d_i \exp i S_{ghost},
\ee
where
\be\label{eq22}
S_{ghost}=\int dx\left[{\bar c}_i\left(F_{ij}\left(H_{jk}+\lambda_l
\frac{\partial H_{jk}}{\partial\phi_l}\right)\right)c_k+
\bar d_i F_{ij} H_{jk} c_k+\bar c_i F_{ij} H_{jk} d_k\right].
\ee
We next insert a factor of 
\be\label{eq23}
\int\dc p_i \dc q_i \exp\frac{-i}{2\alpha}\int dx\left[p_i N_{ij} p_j+
2 p_i K_{ij} q_j\right] {\det}^2 K_{ij}
\ee
into Eq. \eqref{eq4}. The determinant in Eq. \eqref{eq23} leads to
the square of the usual
``Nielsen-Kallosh'' ghost \cite{Nielsen:1978mp,Kallosh:1978de};
we will set $N=K=I$ and henceforth ignore
it. The integrals over $p$ and $q$ in Eq. \eqref{eq23} can be
performed using the $\delta$-functions in Eq. \eqref{eq19} leaving us
with an effective action 
$S_{eff} = S_{cl} + S_{gf} + S_{ghost}$ where
\be\label{eq24}
S_{cl} = \int dx \left(\lc + \lambda_l\frac{\partial\lc}{\partial\phi_l}\right)
\ee
and
\be\label{eq25}
S_{gf} = \int dx\left[-N_i F_{ij}\left(\phi_j+\lambda_j\right)
+\frac{\alpha}{2} N_i N_i - L_i F_{ij} \phi_j + \alpha N_i L_i\right].
\ee
In Eq. \eqref{eq25}, $N_i$ and $L_i$ are ``Nakanish-Lautrup'' fields
used to linearize $S_{gf}$ in the gauge fixing condition of
Eq. \eqref{eq18} and to make it possible to take the limit
$\alpha\rightarrow 0$ \cite{Nakanishi:1966zz,Lautrup:1967zz}.

We now apply these general considerations to the specific instances of
the YM and EH actions.

\section{Yang-Mills Theory}
We now will examine how the LM field can be used to reduce YM theory
to one-loop order. In ref. \cite {McKeon:1992rq} this was considered
using the second order form of the action; here we use the first order
form
\be\label{eq26}
S_{YM} = \int dx\left[-\frac 1 2 F^{a\mu\nu}\left(
\partial_\mu A_\nu^a - \partial_\nu A^a_\mu+g f^{abc} A_\mu^b
A_\nu^c\right) + \frac 1 4 F^{a\mu\nu} F^a_{\mu\nu}\right],
\ee
in which the vector potential $A_\mu^a$ and the field strength
$F^a_{\mu\nu}$ are treated as independent gauge fields $\phi_i$
\cite{McKeon:1994ds}. Using these two field simplifies the interaction
vertices in the theory. Once a LM is introduced for these two fields,
$S_{YM}$ is supplemented by
\be\label{eq27}
S_{LM} = \int dx\left[\frac 1 2 \Lambda^{a\mu\nu}\left(
F^a_{\mu\nu} - f^a_{\mu\nu}\right) + \lambda^{a\nu}
\left(D^{ab\mu} F_{\mu\nu}^b\right)\right],
\ee
where
\begin{subequations}
\be\label{eq28a}
D(A)^{ab}_\mu =\partial_\mu \delta^{ab} + g f^{apb} A^p_\mu
\ee
and
\be\label{eq28b}
D_\mu^{ap} D_\nu^{pb} - D_\nu^{ap} D_\mu^{pb} =
g f^{apb} f_{\mu\nu}^p.
\ee
\end{subequations}
Since $S_{YM}$ in Eq. \eqref{eq26} is invariant under the gauge
transformation 
\begin{subequations}
\be\label{eq29a}
\delta A_\mu^a = D_\mu^{ab} \xi^b 
\ee 
and 
\be\label{eq29b}
\delta F^a_{\mu\nu} = g f^{abc} F_{\mu\nu}^b \xi^c ,
\ee
\end{subequations}
we see that by Eqs. \eqref{eq15} and \eqref{eq17} we have in addition
\begin{subequations}
\be\label{eq30a}
\delta \lambda_\mu^a = g f^{abc} \lambda_{\mu}^b \xi^c ,
\ee  
\be\label{eq30b}
\delta \Lambda^a_{\mu\nu} = g f^{abc} \Lambda_{\mu\nu}^b \xi^c ,
\ee 
\end{subequations}
as well as
\begin{subequations}
\be\label{eq31a}
\delta \lambda_\mu^a = D_\mu^{ab} \zeta^b 
\ee  
\be\label{eq31b}
\delta \Lambda^a_{\mu\nu} = g f^{abc} F_{\mu\nu}^b \zeta^c .
\ee 
\end{subequations}
With the gauge conditions of Eq. \eqref{eq18} taken to be
\be\label{eq32}
\partial\cdot A^a = \partial\cdot\lambda^a = 0
\ee
then $S_{gf}+ S_{ghost}$ of Eqs. \eqref{eq22} and \eqref{eq25} becomes
\be\label{eq33}
\int dx\left[-N^a\partial^\mu\left(A^a_\mu+\lambda^a_\mu\right)
+\frac{\alpha}{2} N^a N^a - L^a\partial^\mu A_\mu^a+\alpha N^aL^a
+{\bar c}^a\partial\cdot D^{ab}(A+\lambda)c^b +
{\bar c}^a\partial\cdot D^{ab}(A) d^b+
{\bar d}^a\partial\cdot D^{ab}(A) c^b\right].
\ee

As $S_{YM}+S_{LM}+S_{gf}+S_{ghost}$ has the structure of Eq. \eqref{eq11},
we see that there are no diagrams beyond one-loop order with the tree level
diagrams being those that occur in normal YM theory and the one-loop
diagrams being doubled. As is demonstrated in Ref. \cite{McKeon:1992rq},
this means that the one-loop divergences that arise can be removed by
an exact renormalization of the coupling $g$ and the wave function
$A_\mu^a$; 
the resulting $\beta$-function associated with the running of $g$ is twice the usual
one-loop $\beta$-function in normal YM theory.

We now turn to how the LM field can be used in conjunction with the EH action.

\section{General Relativity}
We now show how the LM field can be used to limit radiative effects when
used in conjunction with the Palatini (first order) form of the EH action.
The second order form has been considered in Ref. \cite{Brandt:2018lbe}. In
the first order form, only three-point vertices arise which
considerably simplifies the Feynman rules that are needed.

If we use the fields
\begin{subequations}\label{eq34}
  \be\label{eq34a}
  h^{\mu\nu} = \sqrt{-g} g^{\mu\nu}
\ee
\be\label{eq34b}
G_{\mu\nu} ^\lambda = \Gamma_{\mu\nu} ^\lambda  -  \frac 1 2\left(
\delta^\lambda_\mu \Gamma_{\nu\sigma} ^\sigma + \delta^\lambda_\nu \Gamma_{\mu\sigma} ^\sigma
\right),
\ee
\end{subequations}
where $\Gamma_{\mu\nu} ^\lambda$ are the Christoffel symbols,
then the EH action in first order form becomes in $d$ dimensions
\cite{McKeon:2010nf} (with $\kappa^2 = 16\pi G_N$)
\begin{eqnarray}\label{eq35}
  S_{EH} &=&
   \frac{1}{\kappa^2} \int d^d x \,\sqrt{-g}g^{\mu\nu}R_{\mu\nu}(\Gamma)
\nonumber\\
&=&  \frac{1}{\kappa^2} \int d^d x \, h^{\mu\nu}\left(
G_{\mu\nu,\lambda} ^\lambda +\frac{1}{d-1} G_{\lambda\mu} ^\lambda G_{\sigma\nu} ^\sigma
-G_{\sigma\mu} ^\lambda G_{\lambda\nu} ^\sigma
\right).
\end{eqnarray}
This is invariant under the infinitesimal gauge transformations
\begin{subequations}\label{eq36}
  \be\label{eq36a}
\delta h^{\mu\nu} = h^{\mu\rho}\partial_\rho \xi^\nu+h^{\nu\rho}\partial_\rho \xi^\mu
-\partial_\rho\left(h^{\mu\nu}\xi^\rho\right)
\ee
  \be\label{eq36b}
\delta G_{\mu\nu}^\lambda = -\partial_{\mu\nu}^2\xi^\lambda +
\frac 1 2
\left(\delta^\lambda_\mu\partial_\nu+\delta^\lambda_\nu\partial_\mu\right)\partial\cdot\xi
-\xi\cdot\partial G_{\mu\nu}^\lambda+
G_{\mu\nu}^\rho \partial_\rho\xi^\lambda - \left(
G_{\mu\rho}^\lambda \partial_\nu+G_{\nu\rho}^\lambda \partial_\mu
\right)\xi^\rho
\ee
\end{subequations}
which follow from diffeomorphism invariance.

LM fields $\lambda^{\mu\nu}$ and $\Lambda_{\mu\nu}^\lambda$
are associated with the equations
of motion of $h^{\mu\nu}$ and $G^\lambda_{\mu\nu}$, respectively. Again using
Eq. \eqref{eq11}, we see that $S_{EH}$ of Eq. \eqref{eq35} is supplemented by
\begin{eqnarray}\label{eq37}
S_{LM} &=& \int d^d x\bigg[
\lambda^{\mu\nu}\left(
G_{\mu\nu,\lambda} ^\lambda +\frac{1}{d-1} G_{\lambda\mu} ^\lambda G_{\sigma\nu} ^\sigma -
G_{\sigma\mu} ^\lambda G_{\lambda\nu} ^\sigma
\right)  \nonumber \\
&+& \Lambda_{\mu\nu}^\lambda\left(-h^{\mu\nu}_{,\lambda}
+\frac{1}{d-1}
\left(\delta^\mu_\lambda h^{\nu\rho}+\delta^\nu_\lambda h^{\mu\rho}\right)
G_{\rho\sigma}^\sigma
-\left(h^{\mu\rho}G_{\lambda\rho}^\nu+h^{\nu\rho}G_{\lambda\rho}^\mu\right)
\right)\bigg].
\end{eqnarray}

The general gauge invariances of Eqs. \eqref{eq13}, \eqref{eq15} and \eqref{eq17}
that follow from Eq. \eqref{eq36} are
\begin{subequations}
\begin{eqnarray}\label{eq38a}
\delta\lambda^{\mu\nu} = h^{\mu\rho}\partial_\rho\zeta^\nu+
h^{\nu\rho}\partial_\rho\zeta^\mu -\partial_\rho\left(h^{\mu\nu}\zeta^\rho\right)
\nonumber \\
+\lambda^{\mu\rho}\partial_\rho\xi^\nu+\lambda^{\nu\rho}\partial_\rho\xi^\mu
-\partial_\rho\left(\lambda^{\mu\nu}\xi^\rho\right)
\end{eqnarray}
and
\begin{eqnarray}\label{eq38b}
\delta\Lambda_{\mu\nu}^\lambda &=& -\partial_{\mu\nu}^2\zeta^\lambda+
\frac 1 2\left(\delta^\lambda_\mu\partial_\nu+\delta^\lambda_\nu\partial_\mu\right)
\partial\cdot\zeta - \zeta\cdot\partial G_{\mu\nu}^\lambda+
G_{\mu\nu}^\rho\partial_\rho\zeta^\lambda\nonumber\\
&-&\left(G_{\mu\rho}^\lambda\partial_\nu+
G_{\nu\rho}^\lambda\partial_\mu\right)\zeta^\rho
\nonumber\\
&-&\xi\cdot\partial\Lambda_{\mu\nu} ^\lambda+
\Lambda_{\mu\nu}^\rho\partial_\rho\xi^\lambda
-\left(\Lambda_{\mu\rho}^\lambda\partial_\nu+
\Lambda_{\nu\rho}^\lambda\partial_\mu\right)\xi^\rho.
\end{eqnarray}
\end{subequations}
If we choose the gauge fixing conditions \cite{Nishijima:1978wq}
\be\label{eq39}
\partial_\mu h^{\mu\nu} = \partial_\mu\lambda^{\mu\nu} = 0
\ee
then by Eqs. \eqref{eq22} and \eqref{eq25} we find that
\be\label{eq40}
S_{gf} = \frac{1}{\kappa^2} \int d^d x   \left[-N_\nu\partial_\mu\left(
  h^{\mu\nu}+\lambda^{\mu\nu}\right)+\frac \alpha 2 N_\mu N^\mu- L_\nu\partial_\mu h^{\mu\nu}+
  \alpha N_\mu L^\mu
  \right]
\ee
\begin{eqnarray}\label{eq41}
  S_{ghost} &=& \frac{1}{\kappa^2} \int d^d x   \big\{
{\bar c}_\nu\partial_\mu
  \left[\left(h^{\mu\rho}+\lambda^{\mu\rho}\right)\partial_\rho c^\nu +
    \left(h^{\nu\rho}+\lambda^{\nu\rho}\right)\partial_\rho c^\mu
    -\partial_\rho\left(\left(h^{\mu\nu}+\lambda^{\mu\nu}\right)
    c^\rho \right)
    \right]
  \nonumber \\
  &+& {\bar c}_\nu\partial_\mu\left[h^{\mu\rho}\partial_\rho d^\nu +
                                   h^{\nu\rho}\partial_\rho d^\mu
-\partial_\rho\left(h^{\mu\nu} d^\rho\right)\right]
 +           {\bar d}_\nu\partial_\mu\left[h^{\mu\rho}\partial_\rho c^\nu +
                                   h^{\nu\rho}\partial_\rho c^\mu
-\partial_\rho\left(h^{\mu\nu} c^\rho\right)\right]
\big\}
\end{eqnarray}

From Eqs. \eqref{eq35}, \eqref{eq37}, \eqref{eq40} and \eqref{eq41} it is evident that
$\lambda^{\mu\nu}$, $\Lambda_{\mu\nu}^\lambda$, $L_{\mu}$, $d^{\mu}$, ${\bar d}^{\mu}$ all
play the role of a LM fields if one were to simply quantize the EH action of
Eq. \eqref{eq35}. Using the general arguments of section two above, we see that
in any perturbative expansion of radiative effects, there are no contributions beyond
one-loop order with tree effects the same as those coming from the EH action
without the LM contribution and with the one-loop contributions being doubled.

The problem of having to dispose of divergences arising beyond one-loop order consequently
never arises when using the LM field. The one-loop divergences vanish when the
equation of motion are satisfied by the external fields (provided one discards
a surface term) when one considers the EH action by itself. This means that
a shift in the metric field can be used to eliminate these
divergences \cite{tHooft:1974bx,tHooft:2002xp,Buchbinder:1992rb}.
The one-loop divergences that
arise when the metric couples to scalar \cite{tHooft:1974bx},
vector or spinor fields \cite{Deser:1974xq,Deser:1974cy}
no longer are proportional to the equations of motion
and hence cannot be eliminated in this way. However, in all instances the one-loop
divergences are proportional to $R_{\mu\nu}$ and hence they can be absorbed
by $\lambda^{\mu\nu}$ in Eq. \eqref{eq37}. Having the one-loop divergences
absorbed by the LM field when considering the EH action and its
extensions when the metric couples to matter fields is distinct from
what happens when we considered the YM field in the preceding section.
This is discussed in more detail in Ref. \cite{Brandt:2018lbe}.

We now will consider how invariance under BRST transformations can be used when
a LM field is introduced in a classical gauge theory.

\section{BRST Invariance}
The well known BRST (Becchi-Rouet-Stora-Tyutin) \cite{brs74,Tyutin:1975qk,Kugo:1979gm,Nakanishi:1990qm} procedure for
finding a non-linear global Fermionic symmetry for the effective
action in gauge theories can be adapted to the presence
of a LM field. For YM theory, the effective action of Eqs. \eqref{eq26},
\eqref{eq27} and \eqref{eq33} is invariant under the transformation
\begin{subequations}\label{eq42}
  \be\label{eq42a}
\delta A^a_\mu = D_\mu^{ab}(A) c^b\epsilon ,
  \ee
  \be\label{eq42b}
  \delta \lambda^a_\mu = \left(D_\mu^{ab}(A) d^b
  +g f^{apb}\lambda_\mu^{p} c^b \right)\epsilon ,
  \ee
  \be\label{eq42c}
\delta F^a_{\mu\nu} = g f^{abc} F_{\mu\nu}^bc^c\epsilon ,
  \ee
  \be\label{eq42d}
\delta \Lambda^a_{\mu\nu} = g f^{abc} \left(\Lambda_{\mu\nu}^b c^c +
F_{\mu\nu}^b d^c\right) \epsilon ,
\ee      
  \be\label{eq42e}
\delta c^a = \frac 1 2 g f^{abc} c^b c^c\epsilon,\;\;\;
\delta d^a = g f^{abc} c^b d^c\epsilon,
  \ee
  \be\label{eq42g}
\delta N^a = 0 = \delta L^a
  \ee
  \be\label{eq42i}
\delta {\bar c}^a = - N^a\epsilon,\;\;\;
\delta {\bar d}^a = - L^a\epsilon,
\ee      
\end{subequations}
where $\epsilon$ is a constant Grassmann parameter. Upon invoking
the Jacobi identity, it can be verified that for each of the fields
$\phi_i$ in Eq. \eqref{eq42}, $\delta$ is nilpotent so that
\be\label{eq43}
\delta^2\phi_i = 0.
\ee
Since the effective action of Eqs. \eqref{eq26}, \eqref{eq27}
and \eqref{eq33} 
is invariant under this transformation, the argument of
Refs. \cite{Kugo:1979gm,Nakanishi:1990qm,peskin_scroeder}
can be used to establish the
unitarity of first order YM theory when supplemented with a LM field.

Much the same approach can now be used with the first order (Palatini) form
of the EH action when supplemented with a LM field. The BRST transformation
for the second order EH by itself has been considered in
\cite{Nishijima:1978wq,Nakanishi:1990qm,Delbourgo:1976xd,Nakanishi:1977gt,Faizal:2010dw}.
When dealing with the effective action arising from the sum of Eqs. 
\eqref{eq35}, \eqref{eq37}, \eqref{eq40} and \eqref{eq41}, we can use
Eq. \eqref{eq42} as a guide to obtaining the appropriate BRST transformations.
One can establish invariance of the effective action under the
transformations
\begin{subequations}\label{eq44}
\begin{eqnarray}\label{eq44a}
\delta h^{\mu\nu} =\left[h^{\mu\rho}\partial_\rho c^\nu+h^{\nu\rho}\partial_\rho c^\mu
-\partial_\rho\left(h^{\mu\nu} c^\rho\right)\right]\epsilon,
\end{eqnarray}
\begin{eqnarray}\label{eq44b}
\delta \lambda^{\mu\nu} &=&\left[h^{\mu\rho}\partial_\rho d^\nu+h^{\nu\rho}\partial_\rho d^\mu
-\partial_\rho\left(h^{\mu\nu} d^\rho\right)\right]\epsilon \nonumber \\
&+&
\left[\lambda^{\mu\rho}\partial_\rho c^\nu+\lambda^{\nu\rho}\partial_\rho c^\mu
-\partial_\rho\left(\lambda^{\mu\nu} c^\rho\right)\right]\epsilon ,
\end{eqnarray}
\begin{eqnarray}\label{eq44c}
  \delta G_{\mu\nu}^\lambda &=& \big[-\partial^2_{\mu\nu} c^\lambda +
\frac 1 2\left(\delta^\lambda_\mu\partial_\nu+\delta^\lambda_\nu\partial_\mu\right)    
\partial\cdot c- c\cdot\partial G^\lambda_{\mu\nu} \nonumber \\
&+& G_{\mu\nu}^\rho\partial_\rho c^\lambda -
\left(G_{\mu\rho}^\lambda\partial_\nu+G_{\nu\rho}^\lambda\partial_\mu\right)c^\rho\big]\epsilon ,
\end{eqnarray}
\begin{eqnarray}\label{eq44d}
  \delta \Lambda_{\mu\nu}^\lambda &=& \big[-\partial^2_{\mu\nu} d^\lambda +
\frac 1 2\left(\delta^\lambda_\mu\partial_\nu+\delta^\lambda_\nu\partial_\mu\right)    
\partial\cdot d- d\cdot\partial G^\lambda_{\mu\nu} \nonumber \\
&+& G_{\mu\nu}^\rho\partial_\rho d^\lambda -
\left(G_{\mu\rho}^\lambda\partial_\nu+G_{\nu\rho}^\lambda\partial_\mu\right)d^\rho\big]\epsilon 
\nonumber \\
&+&\left[-c\cdot\partial\Lambda_{\mu\nu}^\lambda+\Lambda_{\mu\nu}^\rho\partial_\rho c^\lambda-
\left(\Lambda_{\mu\rho}^\lambda\partial_\nu+\Lambda_{\nu\rho}^\lambda\partial_\mu\right)c^\rho
\right]\epsilon,
\end{eqnarray}
\begin{eqnarray}\label{eq44e}
\delta c^\mu = c^\lambda\partial_\lambda c^\mu\epsilon,
\end{eqnarray}
\begin{eqnarray}\label{eq44f}
  \delta d^\mu = \left(d^\lambda\partial_\lambda c^\mu+
                      c^\lambda\partial_\lambda d^\mu\right)\epsilon,
\end{eqnarray}
\be\label{eq44gh}
\delta N_\mu = 0 = \delta L_\mu,
\ee
\be\label{eq44i}
\delta {\bar c}^\mu = - N^\mu \epsilon ,
\ee
\be\label{eq44j}
\delta {\bar d}^\mu = - L^\mu \epsilon .
\ee
\end{subequations}
If $\phi_i$ is now identified with the fields appearing in Eqs. \eqref{eq44},
we find that once again Eq. \eqref{eq43} is satisfied.

%
%

Greens functions in gauge theories are not all independent because of
the symmetries
present \cite{Slavnov:1972fg,Taylor:1971ff}. Dealing with gauge symmetries through
the BRST identities provides a straightforward way of finding these relations.
In fact, if there is a LM present so that all radiative corrections
are just one-loop with only gauge fields on external legs, the
approach in which the LM field is absent can be easily adopted.
The usual method to derive the BRST equation starts from the symmetry
of the generating functional of connected Green's functions. Then, by
a Legendre transform one obtains the generating functional of the
one-particle irreducible Green's functions. We shall not repeat here
this procedure, which is well known (see, for example,
the chapter 7 of Ref \cite{ryder:book94}). Instead, we will
emphasize the important fact that the BRST equation directly
reflects the symmetry of the gauge theory under BRST transformations.
Using the YM effective action $S_{eff}$ of Eqs.
\eqref{eq26}, \eqref{eq27} and \eqref{eq33} with the BRST invariance of Eq. \eqref{eq42},
we can supplement $S_{eff}$ with source terms \cite{KlubergStern:1974rs}
\be\label{eq45f}
S_s = \int d^d x\left[u^{a}_{\mu}\delta A^{a\mu}+
  {\hat u}^{a}_{\mu}\delta \lambda^{a\mu}+
  w^{a}_{\mu\nu}\delta F^{a\mu\nu}+
  {\hat w}^{a}_{\mu\nu}\delta \Lambda^{a\mu\nu}+
  v^{a} \delta c^{a}+{\hat v}^{a} \delta d^{a}\right]
\ee
where the invariant sources $u_\mu^a$, ${\hat u}_\mu^a$,
$w_{\mu\nu}^a$, ${\hat w}_{\mu\nu}^a$, $v^a$ and ${\hat v}^a$
%
%
are coupled to the non-linear variations of the fields $A^{a\mu}$,
$\lambda^{a \mu}$, $F^{a \mu\nu}$, $\Lambda^{a \mu\nu}$, $c^a$ and $d^a$ 
given by Eq. \eqref{eq42}.
Notice that the variations of ${\bar c}^a$ and ${\bar d}^a$
are linear in the fields
so that the introduction of new sources is not needed here.
Due to the nilpotency \eqref{eq43} of these variations,
we see that similarly to $S_{eff}$, $S_s$ is also invariant
under BRST transformations. Consequently, the total action
\be\label{eq46f}
\Gamma = S_{eff} + S_s
\ee
will also be invariant. This invariance under BRST
transformation \eqref{eq42} can be expressed as
\begin{eqnarray}\label{eq47f}
\int d^d x\bigg[\frac{\delta\Gamma}{\delta A^{a\mu}}{\delta A^{a\mu}}+
\frac{\delta\Gamma}{\delta\lambda^{a\mu}}{\delta \lambda^{a\mu}}+
\frac{\delta\Gamma}{\delta F^{a\mu\nu}}{\delta F^{a\mu\nu}}+          
\frac{\delta\Gamma}{\delta \Lambda^{a\mu\nu}}{\delta \Lambda^{a\mu\nu}}
+\frac{\delta\Gamma}{\delta c^{a}}{\delta c^{a}}+
\frac{\delta\Gamma}{\delta d^{a}}{\delta d^{a}}+
\frac{\delta\Gamma}{\delta {\bar c}^{a}}{\delta {\bar c}^{a}}+
\frac{\delta\Gamma}{\delta {\bar d}^{a}}{\delta {\bar d}^{a}}\bigg]=0.
\end{eqnarray}
Using \eqref{eq45f}, one can write \eqref{eq47f} in the alternative way
\begin{eqnarray}\label{eq48f}
  \int d^d x\bigg[\frac{\delta\Gamma}{\delta A^{a\mu}}
                \frac{\delta\Gamma}{\delta u^{a}_{\mu}}+
\frac{\delta\Gamma}{\delta\lambda^{a\mu}}
\frac{\delta\Gamma}{\delta {\hat u}^{a}_{\mu}}+
\frac{\delta\Gamma}{\delta F^{a\mu\nu}}\frac{\delta\Gamma}{\delta w^{a}_{\mu\nu}}+
\frac{\delta\Gamma}{\delta \Lambda^{a\mu\nu}}
\frac{\delta\Gamma}{\delta {\hat w}^{a}_{\mu\nu}}+
+\frac{\delta\Gamma}{\delta c^{a}}\frac{\delta\Gamma}{\delta v^{a}}+
\frac{\delta\Gamma}{\delta d^{a}}\frac{\delta\Gamma}{\delta {\hat v}^{a}}+
\frac{\delta\Gamma}{\delta {\bar c}^{a}}{\delta {\bar c}^{a}}+
\frac{\delta\Gamma}{\delta {\bar d}^{a}}{\delta {\bar d}^{a}}\bigg]=0.
\end{eqnarray}
To obtain a simpler form of this equation, we note that from \eqref{eq46f}
one can verify the following useful relations involving
$c^a$, ${\bar d}^a$, $u_\mu^a$ and ${\hat u}^a_\mu$
\be\label{eq49f}
\frac{\delta\Gamma}{{\delta\bar c}^a}=-\partial_\mu\left(
\frac{\delta\Gamma}{\delta u^a_\mu}+\frac{\delta\Gamma}{\delta {\hat u}^a_\mu}
\right);\;\;\;
\frac{\delta\Gamma}{{\delta {\bar d}}^a}=-\partial_\mu
\frac{\delta\Gamma}{\delta {u}^a_\mu}.
\ee
With the help of these relations, Eq. \eqref{eq48f} becomes,
after integration by parts
\begin{eqnarray}\label{eq50f}
  \int d^d x\bigg[
    \left(\frac{\delta\Gamma}{\delta A^{a\mu}}
    +\partial_\mu\delta {\bar c}^a +\partial_\mu\delta {\bar d}^a 
    \right)\frac{\delta\Gamma}{\delta u_\mu^a}+
\left(\frac{\delta\Gamma}{\delta\lambda^{a\mu}}
    +\partial_\mu\delta {\bar c}^a \right)\frac{\delta\Gamma}{{\delta\hat u}_\mu^a}
\nonumber \\ +
\frac{\delta\Gamma}{\delta F^{a\mu\nu}}\frac{\delta\Gamma}{\delta w^{a}_{\mu\nu}}+
\frac{\delta\Gamma}{\delta \Lambda^{a\mu\nu}}
\frac{\delta\Gamma}{\delta {\hat w}^{a}_{\mu\nu}}+
+\frac{\delta\Gamma}{\delta c^{a}}\frac{\delta\Gamma}{\delta v^{a}}+
\frac{\delta\Gamma}{\delta d^{a}}\frac{\delta\Gamma}{\delta {\hat v}^{a}}\bigg]=0.
\end{eqnarray}
Finally, defining a new generating functional $\tilde\Gamma$ by
\be\label{eq51f}
\tilde\Gamma = \Gamma - \Gamma_{gf},
\ee
which has the gauge-fixing terms omitted, we obtain from Eq. \eqref{eq50f}
the BRST identities for the effective action $\tilde\Gamma$
\be\label{eq52f}
\int d^d x\bigg[
\frac{\delta\tilde\Gamma}{\delta A^{a\mu}}
\frac{\delta\tilde\Gamma}{\delta u^{a}_{\mu}}+
\frac{\delta\tilde\Gamma}{\delta \lambda^{a\mu}}
\frac{\delta\tilde\Gamma}{\delta {\hat u}^{a}_{\mu}}+
\frac{\delta\tilde\Gamma}{\delta F^{a\mu\nu}}
\frac{\delta\tilde\Gamma}{\delta w^{a}_{\mu\nu}}+
\frac{\delta\tilde\Gamma}{\delta \Lambda^{a\mu\nu}}
\frac{\delta\tilde\Gamma}{\delta {\hat w}^{a}_{\mu\nu}}+
\frac{\delta\tilde\Gamma}{\delta c^{a}}
\frac{\delta\tilde\Gamma}{\delta v^{a}}+
\frac{\delta\tilde\Gamma}{\delta d^{a}}
\frac{\delta\tilde\Gamma}{\delta {\hat v}^{a}}\bigg]=0.
\ee
We note that the above equation is valid to lowest
order. But a similar result can be obtained to higher orders by
considering all fields and sources as being bare quantities \cite{Frenkel:2017xvm}
and replacing the fields by their expectation values
(see Appendix C).
In the standard approaches \cite{taylor:1976b,itzykson:1980b}
such identities have been extensively used to give an all-order inductive
proof of renormalizability. In our case, where graphs beyond one-loop
order are absent, the BRST identities \eqref{eq52f} are also useful
to describe the relations between the proper Green functions
as well as their properties. For example, it follows from \eqref{eq52f}
that the gluon-self energy satisfies the transversality condition
\be\label{eq53f}
\left[
\partial^\mu\frac{\delta^2\tilde\Gamma}{\partial A^{a\mu}\partial A^{b\nu}}
  \right]_{\mbox{fields and sources}=0}   = 0.
\ee
We have verified explicitly this relation to one-loop order.

The BRST transformations of Eq. \eqref{eq44} can be similarly
derived from the effective action arising from the sum
of Eqs. \eqref{eq35}, \eqref{eq37}, \eqref{eq40} and \eqref{eq41}, by
adding the source action (compare with \eqref{eq45f})
\be\label{eq54f}
S_s^{grav} = \int d^d x\left[u_{\mu\nu}\delta h^{\mu\nu}+
  {\hat u}_{\mu\nu}\delta \lambda^{\mu\nu}+
  w^{\mu\nu}_{\rho}\delta G^{\rho}_{\mu\nu}+
 {\hat w}^{\mu\nu}_{\rho}\delta \Lambda^{\rho}_{\mu\nu}+
    v_{\mu} \delta c^{\mu}+{\hat v}_{\mu} \delta d^{\mu}\right],
\ee
where the invariant sources $u_{\mu\nu}$, ${\hat u}_{\mu\nu}$,
$w^{\mu\nu}_\rho$, ${\hat w}^{\mu\nu}_\rho$, $v_\mu$ and ${\hat v}_\mu$
are coupled to the non-linear
variations of $h^{\mu\nu}$, ${\lambda}^{\mu\nu}$,
$G_{\mu\nu}^\rho$, ${\hat w}^{\mu\nu}_\rho$, $c^\mu$ and $d^\mu$.
Proceeding in parallel to the previous analysis, we find that the
BRST equation for the first-order form of the EH action,
when supplemented with a LM field, are given by
(compare with \eqref{eq52f})
\be\label{eq55f}
\int d^d x\bigg[
\frac{\delta\tilde\Gamma}{\delta h^{\mu\nu}}
\frac{\delta\tilde\Gamma}{\delta u_{\mu\nu}}+
\frac{\delta\tilde\Gamma}{\delta \lambda^{\mu\nu}}
\frac{\delta\tilde\Gamma}{\delta {\hat u}_{\mu\nu}}+
\frac{\delta\tilde\Gamma}{\delta G^{\rho}_{\mu\nu}}
\frac{\delta\tilde\Gamma}{\delta w_{\rho}^{\mu\nu}}+
\frac{\delta\tilde\Gamma}{\delta \Lambda^{\rho}_{\mu\nu}}
\frac{\delta\tilde\Gamma}{\delta {\hat w}_{\rho}^{\mu\nu}}+
\frac{\delta\tilde\Gamma}{\delta c^{\mu}}
\frac{\delta\tilde\Gamma}{\delta v_{\mu}}+
\frac{\delta\tilde\Gamma}{\delta d^{\mu}}
\frac{\delta\tilde\Gamma}{\delta {\hat v}_{\mu}}\bigg]=0,
\ee
where the effective action $\tilde\Gamma = S_{eff} + S^{grav}_s$
has the gauge-fixing terms subtracted.

The BRST Eq. \eqref{eq55f} is useful to get basic relations between the proper Green's functions arising in quantum gravity theory with Lagrange multiplier fields. The above identities, which reflect the gauge invariance of the theory, are thus well suited to fix the structure of the counterterms. We have studied the corresponding ultraviolet contributions to one-loop order and verified that these are twice those occurring in standard quantum gravity. But as we have pointed out, there are no higher-order loop contributions in the present theory.  

The above BRST identities yield precise connections between the 
unphysical gauge bosons polarization 
states and the ghost fields. Using the arguments of 
Refs. \cite{Kugo:1979gm,Nakanishi:1990qm,peskin_scroeder}, 
one can show that these relations lead 
to a cancellation of the unphysical degrees of freedom 
in the S-matrix, which establishes the unitarity of the 
Lagrange multiplier theory.

%


\section{Discussion}
In this paper, we have first of all shown that if a standard classical action is
supplemented with a term in which a LM field is used to ensure that
the classical equations of motion are satisfied, then path integral
quantization can be used to show that the usual tree level diagrams are
retained, the usual one-loop diagrams are doubled and all diagrams beyond
one-loop vanish.

We then show how such a LM field can be used when considering the first
order form of the YM and EH actions. The Faddeev-Popov approach is employed to
remove the non-physical degrees of freedom and BRST invariance of the effective actions
in both cases is presented. It is argued that the BRST invariance ensures the
unitarity of these theories when the LM field has been introduced.

We have explicitly verified unitarity to one-loop order
in both YM and EH theories by showing that the imaginary part of the self-energy
amplitudes are proportional to the corresponding cross sections calculated in the
Born approximation. Since all higher loop contributions of amplitudes
are suppressed, unitarity implies that all the contributions to the imaginary
part of the self-energy amplitudes can be
expressed just in terms of the $T$-matrix elements evaluated
in the tree approximation.

The perturbative series in usual quantum field theories may
diverge even asymptotically \cite{Dyson:1952tj}. However, this is not the case
if a LM field is used to suppress higher loop contributions.
In any case, it is generally possible to find a renormalization scheme
in which higher loop calculations only serve to
alter the renormalization group
functions \cite{Chishtie:2015lwk}.

Radiative effects arising beyond one-loop order
have consequences that can be compared with experimental results
when dealing with the Standard Model, so introduction of a LM field
is only of academic interest when dealing with the YM field.
However, the EH action for GR
is non-renormalizable beyond one-loop order and loses all renormalizability
when coupled with matter fields. These deficiencies are overcome if a LM field
is introduced; one could consider this as a way of reconcile quantum
mechanics and General Relativity without having to introduce
extra dimensions, unobserved fields or break unitarity. The LM field
might be seen to play a role analogous to the
Higgs field in the Standard Model in that it serves to make it possible to cope with
ultraviolet divergences. Unfortunately there are currently no experiments
that could test this approach \cite{Donoghue:2017pgk}.

\begin{acknowledgments}
%
F. T. B. and J. F. thank CNPq (Brazil) for financial support. 
D. G. C. M. thanks Roger Macleod for an enlightening discussion 
and FAPESP (Brazil), Grant No. 2018/01073-5, for financial support.
\end{acknowledgments}


\appendix

\section{The Background field}
Background fields \cite{Dewitt:1967ub,Abbott:1980hwabbott82}, have
not been introduced in the body of this paper, but without using a
background field, perturbative calculations are not possible when
dealing with the EH action. This has been viewed as a deficiency of
using this approach to quantize the EH action that we have
employed. However, we wish to demonstrate (using arguments in
Refs. \cite{Culumovic:1989nw,Brandt:2018lbe}) that the background
field is not something introduced in an ad hoc manner and the presence
of background fields is not a deficiency.

We begin with the path integral of Eq. \eqref{eq2} for the generating
functional $Z[j_i]$ for all Feynman diagrams. Connected diagrams are
generated by $W[j_i]$ where \cite{Abbott:1980hwabbott82}
\be\label{a1}
Z[j_i] = \exp \frac{i}{\hbar} W[j_i].
\ee
If we define
\be\label{a2a}
B_i = \frac{\delta W[j_i]}{\delta j_i}
\ee
and make the Legendre transform
\be\label{a2}
\Gamma[B_i] = W[j_i] - \int d^d x B_i j_i
\ee
we obtain $\Gamma[B_i]$, the generating functional for 1PI diagrams.
Eq. \eqref{a1} can now be written as
\be\label{a3}
\exp \frac{i}{\hbar}\Gamma[B_i]   = \int{\cal D}\phi
\exp\frac i \hbar\left[
S[\phi_i] - \int d^d x (\phi_i-B_i)\frac{\delta\Gamma[B_i]}{\delta B_i}\right], 
\ee
as by Eq. \eqref{a2}
\be\label{a4}
j_i = -\frac{\delta\Gamma[B_i]}{\delta B_i}.
\ee
The shift
\be\label{a5}
\phi_i\rightarrow\phi_i+B_i
\ee
results in
\be\label{a6}
\exp \frac{i}{\hbar}\Gamma[B_i]  = \int{\cal D}\phi
\exp\frac i \hbar\left[
S[\phi_i+B_i] - \int d^d x \phi_i\frac{\delta\Gamma[B_i]}{\delta B_i}\right].
\ee
This shows that the background field $B_i$ is not just inserted by
hand; it is the field associated with the current $j_i$ when making a
Legendre transform of the generating functional for connected Green's
functions $W[j_i]$ \cite{McKeon:1985uc,Rebhan:1984bg}.

We can now substitute the expansions
\be\label{a7}
\Gamma[B_i] = \Gamma^{(0)}[B_i]+\hbar \Gamma^{(1)}[B_i]
+\hbar^2 \Gamma^{(2)}[B_i]+\cdots 
\ee
and
\be\label{a8}
{\cal L}(\phi_i+B_i) = {\cal L}(B_i) +\frac{1}{1!}{\cal L}^\prime_{,i}(B_i) \phi_i  
+\frac{1}{2!}{\cal L}^{\prime\prime}_{,ij}(B_i) \phi_i  \phi_j + \cdots  
\ee
(which follow from Eq. \eqref{eq10}) into Eq. \eqref{a3}. This shows
that
\begin{eqnarray}\label{a9}
\exp\frac{i}{\hbar}\left[
\Gamma^{(0)}[B_i]+\hbar \Gamma^{(1)}[B_i]+\hbar^2 \Gamma^{(2)}[B_i]+\cdots 
\right] \nonumber \\
=\exp\left(\frac{i}{\hbar} S[B_i]\right)
\int{\cal D} \phi_i\exp\frac i\hbar\int d^d x\bigg[\left(
\frac{1}{1!}{\cal L}^\prime_{,i}(B_i) \phi_i  
+\frac{1}{2!}{\cal L}^{\prime\prime}_{,ij}(B_i) \phi_i  \phi_j 
+\frac{1}{3!}{\cal L}^{\prime\prime\prime}_{,ijk}(B_i) \phi_i \phi_j \phi_k 
+ \cdots  \right)
\nonumber \\
-\phi_i\left(
\Gamma^{(0)}[B_i]+\hbar \Gamma^{(1)}[B_i]
+\hbar^2 \Gamma^{(2)}[B_i]+\cdots \right)_{, i}
\bigg].
\end{eqnarray}
Upon matching like powers of $\hbar$ in Eq. \eqref{a9}, we see that
\be\label{a10}
\Gamma^{(0)}[B_i] = S[B_i].
\ee
From Eq. \eqref{a10}, it follows that
\be\label{a11}
\int d^d x\left[{\cal L}^\prime_{, i}(B_i) \phi_i -
  \Gamma^{(0)}_{, i}[B_i]\phi_i\right] = 0
\ee
and so the term in Eq. \eqref{a8} that is linear in the quantum field
$\phi_i$ does not contribute to $\Gamma[B_i]$ because of a
cancellation, not because $B_i$ satisfies the classical equations of
motion. Higher orders in $\hbar$ in Eq. \eqref{a9} yield explicit
expressions for the 1PI Green's functions in the loop expansions
\cite{Coleman:1973jx} (though on occasion the loop expansion and an
expansion in powers of $\hbar$ do not match \cite{Holstein:2004dn}).

Having terms linear in $\phi_i$ cancel in \eqref{a9} complicates the
BRST identities to prove renormalizability of gauge 
theories \cite{Frenkel:2018xup,Brandt:2018wxe}.
However, having a
background field can be advantageous in gauge theories, as one
can chose a gauge fixing condition that retains gauge invariance in
the background field. Such a covariant gauge choice in YM theory is
\cite{Honerkamp:1972fd} 
\be\label{a12}
D_\mu^{ab}(B_\mu^a) \phi^{b\mu} =0 ,
\ee
with non-covariant choices considered in
Ref. \cite{McKeon:1985ue}. For the EH action in both first and second
order form, if the background metric is ${\bar g}_{\mu\nu}$ and the
quantum metric $\phi_{\mu\nu}$, then the gauge transformation of
$\phi_{\mu\nu}$ that follows from Eq. \eqref{eq36a} is
\begin{eqnarray}\label{a13}
\delta\phi_{\mu\nu} &=&({\bar g}_{\mu\nu}+\phi_{\mu\nu})_{,\lambda}\theta^\lambda+
({\bar g}_{\mu\lambda}+\phi_{\mu\lambda})\theta^\lambda _{,\nu}+
({\bar g}_{\nu\lambda}+\phi_{\nu\lambda})\theta^\lambda _{,\mu}
\nonumber \\ &\equiv &
\theta_{\mu ;{\bar\nu}}+\theta_{\nu ;{\bar\mu}}+
\theta^\lambda\phi_{\mu\nu;{\bar\lambda}}+
\phi_{\mu\lambda}\theta^\lambda_{;{\bar\nu}}+
\phi_{\nu\lambda}\theta^\lambda_{;{\bar\mu}} \;\;,
\end{eqnarray}
where ``$;\bar\mu$'' denotes a covariant derivative using the
background metric ${\bar g}_{\mu\nu}$.  A suitable gauge choice is
\be\label{a14}
{\phi_{\mu\nu}}^{;\bar\nu} -  k {\phi_\nu}^\nu_{\; ;\bar\mu} = 0
\ee
as this is covariant under transformations in the background
field. The BRST transformation associated with the gauge choice of
Eq. \eqref{a14} is considered in Ref. \cite{Faizal:2010dw}. The use of
Eq. \eqref{a14} with two different choices of the parameter $k$ in
order to have a propagator for $\phi_{\mu\nu}$ that is both transverse
and traceless is considered in Refs. \cite{Brandt:2007td,Brandt:2017oib}.

\section{Supersymmetry}
In this appendix we will show how a LM field can be used in
conjunction with a Fermionic equation of motion. We first will
consider the Wess-Zumino model in which there is a global
supersymmetry, and then briefly examine supergravity.

We use chiral superfields \footnote{We use the conventions
$\eta^{\mu\nu} = {\mbox diag}(+,-,-,-)$, 
$\epsilon^{\alpha\beta} = -i\sigma_2 = \epsilon^{\dot\alpha\dot\beta}=
-\epsilon_{\alpha\beta} = -\epsilon_{\dot\alpha\dot\beta}$,
$\theta^\alpha = \epsilon^{\alpha\beta} \theta_\beta={\theta^{\dot\alpha}}^{\star}$,
$\theta\psi = \theta^\alpha\psi_\alpha = -\theta_\alpha\psi^\alpha$, 
$\bar\theta\bar\psi = \bar\theta_{\dot\alpha}\bar\psi^{\dot\alpha}$,
$\int d^2\theta \theta^\alpha \theta^\beta = -\frac 1 2 \epsilon^{\alpha\beta}$,
$\int d^2\bar\theta \bar\theta_{\dot\alpha} \bar\theta_{\dot\beta} 
= -\frac 1 2 \epsilon_{\dot\alpha\dot\beta}$,
$\left(\sigma^\mu\right)_{\alpha\dot\alpha}=\left(1,\vec\sigma\right) $,
$\left(\bar\sigma^\mu\right)^{\dot\alpha\alpha}=
\left(1,-\vec\sigma\right) $,
$\left(\chi\sigma^\mu\bar\psi\right)^\star =
\left(-\bar\psi{\bar\sigma}^\mu\chi\right)^\star =
\psi\sigma^\mu\bar\chi = -\bar\chi\bar\sigma^\mu\psi$,
$\left\{Q_\alpha,\bar Q_{\dot\beta}\right\}=2\sigma^\mu_{\alpha\dot\beta}
{\bf P}_\mu$,
$Q_\alpha = -i\frac{\partial}{\partial\theta^\alpha}
-i\sigma^\mu_{\alpha\dot\alpha}{\bar\theta} ^{\dot\alpha}{\bf P}_\mu$,
$\bar Q_{\dot\alpha} = i\frac{\partial}{\partial{\bar\theta}^{\dot\alpha}}
+i \theta^\alpha\sigma^\mu_{\alpha\dot\alpha} {\bf P}_\mu$,
${\bf P}_\mu = i\partial_\mu$,
$D_\alpha = \frac{\partial}{\partial\theta^\alpha}
-\sigma^\mu_{\alpha\dot\alpha}{\bar\theta} ^{\dot\alpha}{\bf P}_\mu$
and
$\bar D_{\dot\alpha} = -\frac{\partial}{\partial{\bar\theta}^{\dot\alpha}}
+ \theta^\alpha\sigma^\mu_{\alpha\dot\alpha} {\bf P}_\mu$.}
($\bar D_{\dot{\alpha}}\Phi = 0$)
\be\label{s1}
\Phi(y^\mu,\theta^\alpha) = \varphi(x) 
- i \theta \sigma^\mu\bar\theta\varphi_{,\mu}(x)
-\frac 1 2 (\theta\sigma^\mu\bar\theta) 
(\theta\sigma^\nu\bar\theta)\varphi_{,\mu\nu}(x)
+\sqrt{2}\theta(\psi(x)-i\theta\sigma^\mu\bar\theta\psi_{,\mu}(x))+
\theta\theta F(x);\;\;\; (y^\mu = x^\mu - i\theta\sigma^\mu\bar\theta)
\ee
and $\Lambda$ (with $\varphi\rightarrow\lambda$,
$\psi\rightarrow\chi$, $F\rightarrow G$ in Eq. \eqref{s1}).
The action we consider is $S=S_{WZ} + S_\Lambda$ where 
\be\label{s2}
S_{WZ} = \int d^4 x\left[\int d^2\theta d^2\bar\theta \Phi^\dagger\Phi 
+\int d^2\theta(m\Phi^2+g\Phi^3) 
+\int d^2\bar\theta(m{\Phi^\dagger}^2+g{\Phi^\dagger}^3)\right]
\ee
and
\be\label{s3}
S_\Lambda = \int d^4 x\left[\int d^2\theta d^2\bar\theta
\left(\Lambda^\dagger \Phi + \Phi^\dagger \Lambda\right)
+ \int d^2 \theta\left(2 m \Lambda\Phi+3 g\Lambda\Phi^2\right)
+ \int d^2 \bar\theta\left(2 m \Lambda^\dagger\Phi^\dagger
+3 g\Lambda^\dagger{\Phi^\dagger}^2\right)\right] .
\ee

If we express $S$ in terms of component fields and use the
algebraic equations of motion to eliminate the auxiliary scalar fields
$F$ and $G$, then $S$ is of the form of the usual Wess-Zumino model
for the spinor $\psi$ and the scalar $\varphi$ with $\lambda$ and
$\chi$ acting as LM fields; all radiative effects are hence reduced to
one-loop order as in Eq. \eqref{eq8}.

The arguments given in Ref. \cite{Grisaru:1979wc}
are sufficiently general, so that these can also be used to show that $m \Phi^2+g\Phi^3$
is not renormalized when $S_\Lambda$ is added to $S_{WZ}$ and so we find that the renormalized
quantities ($m_R$, $g_R$, $\Phi_R$) are
related to the bare quantities ($m_0$, $g_0$, $\Phi_0$) by
\be\label{s4}
\Phi_R = Z^{-1/3} \Phi_0,\;\;\; \ m_R = Z^{2/3} m_0,\;\;\; g_R = Z g_0 .
\ee
At one-loop order in momentum space ($\epsilon=2-n/2$, in $n$
dimensions) 
\be\label{s5}
\langle\Phi_0(p)\Phi_0(-p)\rangle = \Phi_0(p) \Phi_0(-p)
\left[1 + g_0^2\left(\frac A\epsilon+A_1\ln{\frac{p^2}{\mu^2}} + \bar A_1
\right)\right],
\ee
which if
\be\label{s6}
\Phi_R = \Phi\left(1+g_0^2\frac{A}{\epsilon}\right)^{1/2}
\ee
then
\be\label{s7}
g_R^2 = \frac{g_0^2}{1+g_0^2\frac A \epsilon}
\ee
For the three-point function we find that
\begin{subequations}\label{s8}
\be \label{s8a}
\langle \Phi_0 \Phi_0 \Phi_0\rangle = 
\Phi_0(p) \Phi_0(q)  \Phi_0(-p-q)  \left[g_0+g_0^3\left(\frac 
    B\epsilon+B_1 L + \bar B_1\right)\right]
\ee
\be\label{s8b}
= \Phi_R(p) \Phi_R(q) \Phi_R(-p-q) 
\left[g_R+g_R^3\left(B_1 L + \bar B_1\right)\right]
\ee
\end{subequations}
provided $A=B$, which is required if $g\Phi^3$ is unrenormalized.

If $g_0$ is to be dimensionless, then we replace $g_0^2$ by
$g_0^2\mu^\epsilon$ where $\mu$ is a dimensionful renormalization mass
scale; Eq.~\eqref{s7} becomes
\be\label{s10}
g_R^2 = \frac{g_0^2\mu^{2\epsilon}}{\left(1+g_0^2 \frac A\epsilon \mu^{2\epsilon}\right)}
\ee 
so that
\be\label{s11}
\mu^2 \frac{\partial g_R^2}{\partial\mu^2} = \epsilon g_R^2 - A g_R^4
=\beta(g_R^2)
\ee
This is an exact expression for the $\beta$-function
with $A$ being twice what occurs in the one-loop $\beta$-function 
when there is no LM field in addition to the WZ field.

Supergravity is a gauge model in which supersymmetry is local. (Three
reviews are in
Refs. \cite{VanNieuwenhuizen:1981ae,Freedman:2012zz,weinberg:book2000}.) As
has been noted, the EH action is renormalizable only at one-loop order
if no matter fields are present. However, supergravity, even if
coupled to matter superfields, is one-loop finite 
\cite{PhysRevLett.38.527}.
Explicit higher loop calculations reveal some unexpected cancellations
of divergences; it is even possible that $N=8$ supergravity is fully
renormalizable due to the presence of some as yet unknown symmetry
\cite{Stelle:2011zeu}.

The introduction of a LM superfield can be used to eliminate the
higher order divergences that potentially appear beyond one-loop order
in a supergravity model. Consider for example $N=1$ supergravity with
fields $\phi_i$ (the graviton, gravitino and auxiliary fields) coupled
to the supersymmetric Standard Model
or Grand Unified model
with fields $\chi_i$ (leptons,
quarks, gauge Bosons and Higgs \cite{Fayet:1976etFayet:1977yc,Sherry:1979tb}). The
two Lagrangians for these are ${\cal L}_{SG}$ and ${\cal L}_{SM}$
respectively, and the total Lagrangian is \cite{Chamseddine:1982jx}
\be\label{s12}
{\cal L}_{T}(\phi_i,\chi_i) = {\cal L}_{SG}(\phi_i) + {\cal L}_{SM}(\phi_i,\chi_i). 
\ee
${\cal L}_{SG}$ is invariant under the local supersymmetry
transformation
\be\label{s13}
\phi_i\rightarrow\phi_i +R_{ij}^I(\phi_i) \xi_j
\ee
and ${\cal L}_T$ is invariant under
\begin{subequations}\label{s14}
\be\label{s14a}
\phi_i\rightarrow\phi_i +R_{ij}^T(\phi_i, \chi_i)]\xi_j 
\ee
and
\be\label{s14b}
\chi_i\rightarrow\phi_i +S_{ij}^I(\phi_i,\chi_i)\xi_j  ,
\ee
\end{subequations}
where in \eqref{s14a} 
\be\label{s16}
R_{ij}^T(\phi_i\chi_i) = R_{ij}^I(\phi_i)+ R_{ij}^{II}(\phi_i,\chi_i) . 
\ee
We now consider the action
\be\label{s17}
S = \int d^d x\int d\theta\left(
{\cal L}_T + \lambda_i\frac{\partial{\cal L}_{SG}}{\partial \phi_i}.
\right)
\ee
The LM superfield $\lambda_i$ in Eq. \eqref{s17} is only associated
with the supergravity Lagrangian${\cal L}_{SG}(\phi_i)$. The general
arguments following Eq. \eqref{eq12} show that the vierbein field 
$e_\mu^i$ and the gravitino field $\psi_\mu$ have no propagator though
there is a propagator in which these fields mix with their associated
LM fields. All the matter fields $\chi_i$ in ${\cal L}_{SM}$ have
their own propagator. Vertices arising from ${\cal L}_{SG}$ have at
most one LM field; $\phi_i$ can occur multiple times on vertices
arising from ${\cal L}_{SG}$  and ${\cal L}_{SM}$ while $\chi_i$
appears only on vertices in ${\cal L}_{SM}$. 

We see that all diagrams with virtual contributions  involving
$e_\mu^i$ or $\psi_\mu$ are restricted to one-loop order and are twice
those coming from ${\cal L}_{SG}$ alone. Virtual matter fields can
occur in higher order diagrams in the loop expansion. External fields
$e_\mu^i$ and $\psi_\mu$ can occur on both the one-loop diagrams
coming from ${\cal L}_{SG}$ alone and diagrams at any order in the
loop expansion coming from ${\cal L}_{SM}$.
All one-loop subdiagrams coming from higher order diagrams with
external $e^i_\mu$ and $\psi_\mu$
are finite   \cite{PhysRevLett.38.527}
and so we can infer that these higher order diagrams are themselves
finite. Divergences in diagrams which involve matter fields (either
externally or an internal lime) can be removed through
renormalization. 

Quantization can proceed using the Faddeev-Popov procedure as was
employed in Eqs. \eqref{eq18}, \eqref{eq19}, \eqref{eq20},
\eqref{eq21}, \eqref{eq22}, \eqref{eq23}, \eqref{eq24} and \eqref{eq25}.

\section{Slavnov-Taylor identities}
In this appendix we will outline the standard derivation of 
Slavnov-Taylor identities \cite{{Slavnov:1972fg,Taylor:1971ff}}.
Let us consider the generating functional of \eqref{eq2}
\be\label{b1}
Z[j] = \int{\cal D}\phi \exp i\left( S[\phi] +  j_i\phi_i\right) .
\ee
(Here ``i'' may be a continuous or discrete index; we use De Witt's extension
of the Einstein summation convention. We also consider units such that
$\hbar=1$.)
One can always perform a field transformation such that
\be\label{b2}
\phi_i\rightarrow\phi_i + \epsilon F_i[\phi],
\ee
where $F_i[\phi]$ is a general nonlinear functional of the fields.
Then, up to first order in $\epsilon$, we obtain from Eq. \eqref{b1}
\be\label{b3}
Z[j] = Z[j] + \epsilon \int{\cal D}\phi\left[i\left(
  \frac{\delta S}{\delta\phi_i} + j_i\right)F_i[\phi] -
  \frac{\delta F_i}{\delta\phi_i}\right]\exp i\left( S[\phi] +  j_i\phi_i\right) .
\ee
Replacing the $\phi$ dependence inside the square bracket of
\eqref{b3} by $\frac{1}{i}\frac{\delta}{\delta j}$
acting on $Z[j]$, we obtain
\be\label{b4}
\left\{i\left(\frac{\delta S}{\delta\phi_i}
  \left[\frac 1 i\frac{\delta}{\delta j}\right]
  +j_i\right) F_i\left[\frac{1}{i}\frac{\delta}{\delta j}\right]
-\frac{\delta F_i}{\delta\phi_i}\left[\frac 1 i\frac{\delta}{\delta j}\right]
\right\}Z[j] = 0.
\ee
The identity in Eq. \eqref{b4} is a direct consequence of the independence
of $Z[j]$ on field redefinitions. The simplest case occurs when $F_i[\phi]$ is
independent of $\phi$ so that the transformation is simply a field independent
translation. (In this case we have the Dyson-Schwinger equation.)
Here we are interested in the general case when there is a
symmetry in the classical action, so that
\be\label{b5}
\frac{\delta S}{\delta \phi_i} F_i[\phi_j] = 0;
\ee
we also assume that the transformations do not change the path integral measure
so that
\be\label{b6}
\frac{\delta F_i[\phi_j]}{\delta\phi_i} = 0.
\ee
For these symmetry transformations, we obtain from Eq. \eqref{b4}
\be\label{b7}
j_i F_i \left[\frac 1 i\frac{\delta}{\delta j}\right] Z[j] = 0.
\ee
Using
\be\label{b8}
j_i = -\frac{\delta \Gamma[\langle\phi\rangle]}{\delta \langle\phi_i\rangle},
\ee
where $\Gamma[\langle\phi\rangle]$ is
the generator of the 1PI Green's functions, Eq. \eqref{b7}
can be written as
\be\label{b9}
\frac{\delta \Gamma[\langle\phi\rangle]}{\delta \langle\phi_i\rangle}
F_i \left[\frac 1 i\frac{\delta}{\delta j}\right] Z[j] = 0.
\ee
Since
\be\label{b10}
\frac{F_i \left[\frac 1 i\frac{\delta}{\delta j}\right] Z[j] }
{Z[j]}  = \langle F_i[\phi]\rangle_j ,
\ee
Eq. \eqref{b9} can be written as
\be\label{b11}
\frac{\delta \Gamma[\langle\phi\rangle]}{\delta \langle\phi_i\rangle}
\langle F_i[\phi]\rangle_j = 0.
\ee
Eq. \eqref{b11} shows that $\Gamma$ is invariant under
\be\label{b12}
\langle \phi_i \rangle \rightarrow \langle \phi_i\rangle
+ \epsilon \langle F_i[\phi]\rangle
\ee
(compare with \eqref{b2}). In general, when $F[\phi]$ is non-linear,
$\langle F_i[\phi]\rangle \ne F_i[\langle\phi\rangle]$.

Let us now add some extra source terms to $S[\phi]$ in such a way that
the generating functional becomes
\be\label{b13}
Z[j,K] =
\int{\cal D}\phi \exp i\left( S[\phi] +  j_i\phi_i + F_i K_i\right) ,
= \exp i W[j,K]
\ee
where the invariant sources $K_i$ are coupled to the 
non-linear variations $F_i$ of the fields. In the specific examples of 
the gauge theories of gravity and Yang-Mills (see section V), 
this excludes the variations of the anti-ghosts, which are linear.
But it may be shown that such variations are cancelled by the
corresponding variations
of the gauge-fixing terms. Using this property, we will leave out
the gauge-fixing terms as well as such 
variations and consider in what 
follows only the fields whose variations are non-linear.
Using again the symmetry of $S[\phi]$ and the invariance of the path integral
measure, we obtain from \eqref{b13}
\be\label{b14}
i\left[{\delta F_i} K_i + 
  j_i F_i \left[\frac 1 i\frac{\delta}{\delta j}\right]
\right] Z[j,K] = 0.
\ee
Considering the class of transformations which are nilpotent, so that
$\delta F_i = 0$ and using 
\be\label{b15}
j_i = -\frac{\delta \Gamma[\langle\phi\rangle,K]}{\delta \langle\phi_i\rangle},
\ee
we obtain from Eq. \eqref{b14}
\be\label{b16}
\langle F_i\rangle_{j,K} \frac{\delta \Gamma[\langle\phi\rangle, K]}
        {\delta\langle\phi_i\rangle} = 0.
\ee
Finally, from the Legendre transformation
\be\label{b17}
\Gamma[\langle\phi\rangle,K] = W[j,K] - \langle\phi_i\rangle j_i
\ee
we have
\be\label{b18}
\frac{\delta\Gamma[\langle\phi\rangle,K]}{\delta K_i}
= \frac{\delta W[j,K]}{\delta K_i} = \langle F_i \rangle_{j,K}
\ee
so that Eq. \eqref{b16} can be written as    
\be\label{b19}
\frac{\delta\Gamma[\langle\phi\rangle,K]}{\delta K_i}
\frac{\delta \Gamma[\langle\phi\rangle, K]}
{\delta\langle\phi_i\rangle} = 0.
\ee
This equation can be used to find relations between Green's functions that  
follow from gauge invariance, such as those of Eqs. \eqref{eq52f} and \eqref{eq55f}  
obtained in the Yang-Mills theory and general relativity, respectively.




\end{document}